\documentstyle[osa,eqsecnum,manuscript]{revtex}
\tightenlines
\begin{document}

\title{Quantum noise in optical fibers II: \\
Raman jitter in soliton communications}

\author{J. F. Corney$^{1,2}$ and P. D. Drummond$^{1}$}

\address{$^{1}$Department of Physics, The University of Queensland, St. Lucia, QLD 4072, Australia \\
$^{2}$Department of Mathematical Modelling, Technical University of Denmark, DK-2800 Lyngby, Denmark}

\date{\today{}}

\maketitle
\begin{abstract}

\noindent The dynamics of a soliton propagating in a single-mode optical fiber
with gain, loss, and  Raman coupling to thermal phonons is analyzed. Using both soliton perturbation
theory and exact numerical techniques, we predict that intrinsic thermal quantum
noise from the phonon reservoirs is a larger source of jitter and
other perturbations than the gain-related Gordon-Haus noise,  for short pulses (\( \lesssim 1ps \)), assuming typical fiber parameters. The size of the
Raman timing jitter is evaluated for both bright and dark (topological) solitons,
and is larger for bright solitons. Because Raman thermal quantum noise is a
nonlinear, multiplicative noise source, these effects are stronger for the more
intense pulses needed to propagate as solitons in the short-pulse regime. Thus Raman
noise may place additional limitations on fiber-optical communications and networking
using ultrafast (subpicosecond) pulses.
\end{abstract}


\section{Introduction}

In this paper, we analyze in some detail the effects of Raman noise on solitons.
In particular we derive approximate analytic expressions and provide further
detail for the precise numerical results published earlier\cite{s147}. The
motivation for this study is essentially that coupling to phonons is one property
of a solid medium that definitely does not obey the nonlinear Schr\"{o}dinger
equation. The presence of Raman interactions plays a major role in perturbing the fundamental
soliton behaviour of the nonlinear Schr\"{o}dinger equation in optical fibers.  This perturbation is in addition to the more straightforward gain/loss effects that produce the well-known Gordon-Haus effect\cite{s35}.

The complete derivation of the quantum theory for optical fibers is given in
an earlier paper\cite{DruCor99a}, denoted (QNI). That paper presented a detailed derivation of the quantum Hamiltonian, and included quantum noise effects due to nonlinearities,
gain, loss, Raman reservoirs and Brillouin scattering.  Phase-space techniques allowed the quantum Heisenberg equations of motion to be mapped onto stochastic partial differential equations.  The result was a generalized nonlinear Schr\"odinger equation, which can be solved numerically or with perturbative analytical techniques.

The starting point for this paper is the phase-space equation for the case of a single polarization mode, obtained using a truncated Wigner representation, which is accurate in the limit of large
photon number. We use both soliton perturbation theory and numerical integration of
the phase-space equation to calculate effects on soliton propagation of all known quantum noise sources, with good agreement between the two methods.

Our main result is that the Raman noise due to thermal phonon reservoirs is
strongly dependent on both temperature and pulse intensity. At room temperature,
this means that Raman jitter and phase noise become steadily more important
as the pulse intensity is increased, which occurs when a shorter soliton pulse
is required for a given fiber dispersion. Using typical fiber parameters, we
estimate that Raman-induced jitter is more important than the well-known Gordon-Haus jitter for pulses shorter than about one picosecond. Although we do not analyze
this in detail here, we note that similar perturbations may occur during the
collision of short pulses in a frequency-multiplexed environment.

\section{Raman-Schr\"{o}dinger Model}
\label{PT}
We begin with the Raman-modified stochastic nonlinear Schr\"odinger equation [Eq.\ (6.3) of (QNI)], obtained using the Wigner representation, for simplicity:

\begin{eqnarray}
\frac{\partial }{\partial \zeta }\phi (\tau ,\zeta ) & = & -\int ^{\infty }_{0 }d\tau' g(\tau -\tau' )\phi (\tau' ,\zeta )+\Gamma (\tau ,\zeta )+\nonumber  \\
 & + & i\left[\pm \frac{1}{2}\frac{\partial ^{2}\phi }{\partial \tau ^{2}}+\int ^{\infty }_{0 }d\tau' h(\tau -\tau' )\phi ^{*}(\tau' ,\zeta )\phi (\tau' ,\zeta )+i\Gamma ^{R}(\tau ,\zeta )\right]\phi (\tau ,\zeta ) \, \, .
\label{RMS}
\end{eqnarray}
Here \( \phi =\Psi \sqrt{vt_{0}/\overline{n}} \) is a dimensionless photon
field amplitude, while $\tau =(t-x/v)/t_{0}$ and
$\zeta =x/x_{0}$, where \( t_{0} \) is a typical pulse duration used for
scaling purposes and $ x_{0}=t_{0}^{2}/|k''|$ is a characteristic dispersion length.  The group velocity $v$ and the dispersion relation $k''$ are calculated at the carrier frequency $\omega_0$. 

Apart from a cut-off dependent vacuum noise,
the photon flux is \({ \cal J}= |\phi |^{2}\overline{n}/t_{0} \), where
\( \overline{n}=|k''|{\cal A}c/(n_{2}\hbar \omega _{0}^{2}t_{0})=v^{2}t_{0}/\chi x_{0} \)
is the typical number of photons in a soliton pulse of width \( t_{0} \), again
for scaling purposes. In this definition, the fibre  is assumed to have a modal cross-sectional area ${\cal A}$ and a change in refractive index per unit intensity of $n_2$.  The
positive sign in front of the second derivative term applies for anomalous dispersion (\( k''<0 \)),  and the negative sign applies for normal dispersion (\( k''>0 \)). The functions $g$ and  $h$ are gain/loss and Raman scattering response functions respectively, while $\Gamma$ and $\Gamma^R$ are stochastic terms, discussed below.

  Similar, but more accurate, equations occur with the positive-$P$ representation, although in this case, the phase-space dimension is doubled.  In order to simplify the calculations further, we assume that gain and loss in
the fiber are broadband relative to the soliton bandwidth, and balance exactly. This requires that the amplifier
sections in the fiber are sufficiently close together (of the order of the soliton scaling length or
less) so that the soliton can propagate without distortion\cite{e375}. 

For the analytic calculations, we also assume that the Raman nonlinear response function is instantaneous
on the timescale of the soliton width. This is equivalent to assuming that
the phonon modes are heavily damped, and means that
the Raman coupling leads to only incoherent scattering of the propagating radiation.
While this approximation neglects the well-known self-frequency shift\cite{MooWonHau94,AtiMysChrGal,s11}, we find that the self-frequency shift by itself is not a major cause of jitter for the distance scales we consider here. This assumption can be improved at the expense of
more complicated analytic calculations. However, the full equations are used in the
numerical simulations, which agree quite well with our analytic predictions.

The Raman-modified equation then reduces to 
\begin{eqnarray}
\frac{\partial }{\partial \zeta }\phi (\tau ,\zeta )=\left[ \pm \frac{i}{2}\frac{\partial ^{2}}{\partial \tau ^{2}}+i\phi ^{*}(\tau ,\zeta )\phi (\tau ,\zeta )\right] \phi (\tau ,\zeta )+\Gamma^{C} (\tau ,\zeta )\, \, , & \label{NLSE} 
\end{eqnarray}
 where the bracketed term represents the usual nonlinear Schr\"odinger (NLS) equation in normalized, propagative form. The combined noise sources have been grouped together
as 
\begin{eqnarray}
\Gamma^{C} (\tau ,\zeta )=\Gamma (\tau ,\zeta ) +i\Gamma ^{R}(\tau ,\zeta )\phi (\tau ,\zeta )\, \, . & 
\end{eqnarray}

\subsection{Initial conditions and quantum evolution}

Equation (\ref{NLSE}) is a complex-number equation that can accurately represent quantum operator evolution through the inclusion of various noise sources.  In the absence of any noise sources, this equation reduces to the classical nonlinear Schr\"odinger equation.  This deterministic limit corresponds to taking $\overline{n} \rightarrow \infty$.  As well as the noise sources explicitly appearing in  Eq. (\ref{NLSE}), there must be noise in the initial conditions to properly represent a quantum state in the Wigner representation.  Regardless of the initial quantum state chosen, there must be at least minimal level of initial fluctuations in $\phi$ to satisfy Heisenberg's uncertainty principle.  We choose to begin with a multimode coherent state, which contains this minimal level of initial quantum noise and which is an accurate model of mode-locked laser output.  This is also the simplest model for the output of mode-locked lasers, and we note that, in general, there could be extra technical noise.  For coherent inputs, the Wigner vacuum fluctuations are Gaussian, and are correlated as 
\begin{eqnarray}
\langle \Delta \phi(\tau ,0)\Delta \phi^{*}(\tau' ,0)\rangle  & = & \frac{1}{2\overline{n}}\delta (\tau -\tau' ).
\end{eqnarray}

Physical quantities can be calculated from this phase-space simulation by averaging products of  $\phi$ and $\phi^*$ over many stochastic trajectories.  In this Wigner representation, these stochastic averages correspond to the ensemble averages of symmetrically-ordered products of quantum operators, such as those representing homodyne measurements and other measurements of phase. 

\subsection{Wigner noise}

Both fiber loss and the presence of a gain medium each contribute quantum noise
to the equations in the symmetrically-ordered Wigner representation. The complex gain/absorption
noise enters the nonlinear Sch\"odinger (NLS) equation through an additive stochastic term \( \Gamma \), whose correlations are:
\begin{eqnarray}
\langle \Gamma(\Omega ,\zeta )\Gamma ^{*}(\Omega' ,\zeta' )\rangle =\frac{(\alpha ^{G}+\alpha ^{A})}{2\overline{n}}\delta (\zeta -\zeta' )\delta (\Omega +\Omega' ), & \label{gain_{c}or} 
\end{eqnarray}
 where \( \Gamma(\Omega ,\zeta ) \) is the Fourier transform of the noise
source: 
\begin{eqnarray}
\Gamma(\Omega ,\zeta )=\frac{1}{\sqrt{2\pi }}\int ^{\infty }_{-\infty }d\tau \Gamma(\tau ,\zeta )\exp(i\Omega \tau).
\end{eqnarray} 
The dimensionless intensity gain and loss are given by $\alpha^{G}$ and $\alpha ^{A}$, respectively.

Similarly, the real Raman noise, which appears as a multiplicative stochastic
variable \( \Gamma ^{R} \), has correlations 
\begin{eqnarray}
\langle \Gamma ^{R}(\Omega ,\zeta )\Gamma ^{R}(\Omega' ,\zeta' )\rangle =\frac{1}{\overline{n}}\delta (\zeta -\zeta' )\delta (\Omega +\Omega' )\left[ n_{th}(\Omega )+\frac{1}{2}\right] \alpha ^{R}({\Omega }), & \label{Raman_{c}or} 
\end{eqnarray}
 where the thermal Bose distribution is given by \( n_{th}(\Omega )=\left[ \exp {(\hbar |\Omega |/k_BTt _{0})}-1\right] ^{-1} \) and where $\alpha ^{R}({\Omega })$ is the Raman gain, whose profile is given in Fig.~1 of (QNI).
Thus the Raman noise is strongly temperature dependent, but it also contains
a spontaneous component which provides vacuum fluctuations even at \( T=0 \).  

As the $\overline{n}$ dependence of all the noise correlations show, the classical limit of these quantum calculations is the deterministic nonlinear Schr\"odinger equation.  The problem of jitter in soliton communications is an example of how intrinsic quantum features can have a direct macroscopic consequence, even in a way that impinges on current developments of applied technology.  There are, of course, classical contributions to jitter, such as noise arising from technical sources.  However, it is the jitter contributions from essentially quantum processes, namely spontaneous emission in fibre amplifiers, that is the current limiting factor in soliton based communications systems.  Other jitter calculations rely on a classical formulation with an empirical addition of amplifier noise, and important predictions of the Gordon-Haus effect have been obtained.  Nevertheless, this quantum treatment presented here of all known noise sources is necessary to determine the limiting effects of other intrinsic noise sources, which become important for shorter pulses and longer dispersion lengths.    

In the absence of the noise sources, the phase-space equations have stationary solutions
in the form of bright (\( + \)) or dark (\( - \)) solitons.
Solitons are solitary waves in which the effects of dispersion are balanced
by nonlinear effects, to produce a stationary pulse that is robust in the presence
of perturbations. We note here that, in reality, the Raman response function is noninstantaneous, which causes a redshift in the soliton frequency.  This soliton self-frequency shift is a deterministic effect, and
so can be neglected in the treatment of noise effects, to a first approximation.
The accuracy of this approximation will be evident in the subsequent comparison
of analytic with numerical results. The numerical results all include the complete
nonlinear response function, rather than the approximate instantaneous form
given above.  

Excessive self-frequency shift may cause problems when finite bandwidth elements are used.  However, it has been shown\cite{SFS_comp} that bandwidth-limited gain can in fact cancel the effect of the Raman redshift, by pulling the soliton back towards the centre of the spectral band.  In the simulations we show in this paper, the total redshift is estimated to be $\Delta f \simeq 0.02 {\rm THz}$, which is small compared to the total width of the gain spectrum in typical fiber laser amplifiers ($\Delta \nu \simeq 3 {\rm THz}$)\cite{s05}.

\section{Perturbation Theory}

We now proceed to derive the approximate analytic expressions for the effects
of noise on soliton jitter, using soliton perturbation theory\cite{s24,s53,s55,s58,s65,s83}, for both bright and dark solitons.

\subsection{Bright solitons}

The stationary soliton of Eq.\ (\ref{NLSE}) for anomalous dispersion is: 
\begin{eqnarray}
\phi _{\textrm{bright}}(\tau ,\zeta )=A{\textrm{sech}}[A\tau -q(\zeta )]\exp[iV\tau +i\theta (\zeta )],\label{bright_{s}ol} 
\end{eqnarray}
 where \( \partial q/\partial \zeta =VA \) and \( \partial \theta /\partial \zeta =(A^{2}-V^{2})/2 \),
with amplitude A and velocity V. Following the method presented by Haus et al\cite{s53,s55},
we treat the effects of the noise terms as perturbations around a soliton solution
whose parameters vary slowly with \( \zeta  \): 
\begin{eqnarray}
\phi (\tau ,\zeta )=\overline{\phi }(\tau ,\zeta )+\Delta \phi (\tau ,\zeta ),\label{pert} 
\end{eqnarray}
 where the unperturbed soliton solution is given by: 
\begin{eqnarray}
\overline{\phi }(\tau ,\zeta )=A(\zeta ){\textrm{sech}}[A(\zeta )\tau -q(\zeta )]\exp[iV(\zeta )\tau +i\theta (\zeta )] & 
\end{eqnarray}
 for a bright soliton. Substituting Eq.\ (\ref{pert}) into Eq.\ (\ref{NLSE})
gives the following linearized equation (first order in \( \Delta \phi (\tau ,\zeta ) \))
\begin{eqnarray}
\frac{\partial }{\partial \zeta }\Delta \phi (\tau ,\zeta )=\left[ \pm \frac{i}{2}\frac{\partial ^{2}}{\partial \tau ^{2}}+i2\overline{\phi}^{*}(\tau ,\zeta )\overline{\phi }(\tau ,\zeta )\right] \Delta \phi (\tau ,\zeta )+i\overline{\phi }(\tau ,\zeta )^{2}\Delta \phi ^{*}(\tau ,\zeta )+\overline{\Gamma }(\tau ,\zeta )\, \, , & \label{linear} 
\end{eqnarray}
 where the linearized noise source \( \overline{\Gamma }(\tau ,\zeta ) \) is
defined as: 
\begin{eqnarray}
\overline{\Gamma }(\tau ,\zeta )=\Gamma(\tau ,\zeta ) +i\Gamma ^{R}(\tau ,\zeta )\overline{\phi }(\tau ,\zeta )\, \, . & 
\end{eqnarray}

Now we wish to determine the evolution of the soliton parameters as a function
of propagation distance \( \zeta  \). To do this, we expand the perturbation
in terms of the soliton parameters plus a continuum term: 
\begin{eqnarray}
\Delta \phi (\tau ,\zeta ) & = & \sum _{i}\frac{\partial \overline{\phi }(\tau ,\zeta )}{\partial P_{i}}\Delta P_{i}+\Delta \phi _{c}(\tau ,\zeta )\nonumber \\
 & = & \sum _{i}f_{P_{i}}\Delta P_{i}+\Delta \phi _{c}(\tau ,\zeta ).\label{expan} 
\end{eqnarray}
 where \( P_{i}\in \{V,q,A,\theta \} \). The projection functions for each
parameter are 
\begin{eqnarray}
f_{A} & = & \left[ \frac{1}{A}-\tau \tanh (A\tau -q)\right] \overline{\phi },\nonumber \\
f_{q} & = & \tanh (A\tau -q)\overline{\phi },\nonumber \\
f_{V} & = & i\tau \overline{\phi },\nonumber \\
f_{\theta } & = & i\overline{\phi }.
\end{eqnarray}
Since the linearised equation Eq.\ (\ref{linear}) is not self-adjoint, these eigenfunctions are not orthogonal.  In order to select out the evolution of particular parameters, we therefore choose an alternative set of functions:
\begin{eqnarray}
\underline{f_{A}} & = &\overline{\phi },\nonumber \\
\underline{f_{q}} & = & \tau \overline{\phi },\nonumber \\
\underline{f_{V}} & = & i\tanh (A\tau -q)\overline{\phi },\nonumber \\
\underline{f_{\theta }} & = & i\tau \tanh (A\tau -q)\overline{\phi }.
\end{eqnarray}
These are the eigenfunctions of the adjoint equation Eq.\ (\ref{linear}), and obey the orthogonality
condition 
\begin{eqnarray}
\Re \left\{ \int _{-\infty }^{\infty }d\tau f_{P_{i}}\underline{f_{P_{j}}}^{*}\right\} =\delta _{i,j}.
\end{eqnarray}
Substituting the Taylor expansion [Eq.\ (\ref{expan})] into the linearized
equation [Eq.\ (\ref{linear})] and using the functions \( \underline{f_{P_{i}}} \) to project out
particular parameters shows that the growth of fluctuations in position $\Delta q$ is governed by
\begin{eqnarray}
\frac{\partial }{\partial \zeta }\Delta q(\zeta ) &=& A\Delta V(\zeta ) + \Gamma_q(\zeta ) \nonumber \\
\frac{\partial }{\partial \zeta }\Delta V(\zeta ) &=& \Gamma_V(\zeta ),
\end{eqnarray}
where we have taken the unperturbed velocity to be zero: $V = 0$.  The stochastic terms are defined as
\begin{eqnarray}\Gamma _{P_{i}}(\zeta )=\Re \left\{ \int _{-\infty }^{\infty }d\tau \underline{f_{p}}^{*}(\zeta )\overline{\Gamma }(\tau ,\zeta )\right\} . & \label{sp} 
\end{eqnarray}
Here we have assumed that the perturbations in the continuum $\phi _{c}$ are orthogonal to the  \( \underline{f_{P_{i}}} \).  This depends on such perturbations dispersing sufficiently rapidly away from the region around the soliton.  In fact, any nonsoliton perturbation will disperse and would also move away from the soliton, since the group velocity for any linear perturbations will be different to the propagation velocity of the  soliton.  

We wish to find the growth of fluctuations in position \( q(\zeta ) \). Because
the position depends on the soliton frequency \( V \), the contributions arising
from both \( \Gamma _{q} \) and \( \Gamma _{V} \) must be considered. Firstly,
\begin{eqnarray}
\Gamma _{q}(\zeta ) & = & \Re \left\{ \int _{-\infty }^{\infty }d\tau A\tau {\textrm{sech}}(A\tau -q)\exp(-iV\tau -i\theta )\overline{\Gamma} (\tau ,\zeta )\right\} \nonumber \\
 & = & \int _{-\infty }^{\infty }d\tau A\tau {\textrm{sech}}(A\tau -q)\Re \{\exp(-iV\tau -i\theta )\Gamma \},
\end{eqnarray}
 and 
\begin{eqnarray}
\Gamma _{V}(\zeta ) & = & \Re \left\{ \int _{-\infty }^{\infty }d\tau A (-i){\textrm{sech}}(A\tau -q)\tanh (A\tau -q)\exp(-iV\tau -i\theta )\overline{\Gamma} (\tau ,\zeta )\right\} \nonumber \\
 & = & \int _{-\infty }^{\infty }d\tau A{\textrm{sech}}(A\tau -q)\tanh (A\tau -q)\left[ A{\textrm{sech}}(A\tau -q)\Gamma ^{R}+\Im \{\exp(-iV\tau -i\theta )\Gamma \}\right] .
\end{eqnarray}
From this we can calculate the growth of the fluctuations in velocity: 
\begin{eqnarray}
\Delta V(\zeta ) & = & \Delta V(0)+\int _{0}^{\zeta }d\zeta' \Gamma _{V}(\zeta ')\nonumber \\
 & = & \Re \left\{ \int _{-\infty }^{\infty }\Delta \phi (\tau ,\zeta )\underline{f_{V}}^{*}d\tau \right\} +\int _{0}^{\zeta }d\zeta' \Gamma _{V}(\zeta ').\label{delV} 
\end{eqnarray}
 Using the noise correlations calculated above, the correlations in the velocity
fluctuations can now be calculated: 
\begin{eqnarray}
\langle \Delta V(\zeta )\Delta V^{*}(\zeta' )\rangle  & = & \bigl <{\Delta V(0)\Delta V^{*}(0)}\bigr >+\int _{0}^{\zeta }\int _{0}^{\zeta' }d\zeta''d \zeta''' \langle \Gamma _{V}(\zeta'' )\Gamma _{V}^{*}(\zeta''' )\rangle \nonumber \\
 & = & \frac{A}{6\overline{n}}+\left[ \frac{\alpha ^{G}A}{3\overline{n}}+\frac{2A^2{\mathcal{I}}(t_{0})}{\overline{n}}\right] \zeta \quad \quad \zeta <\zeta' ,
\end{eqnarray}
 where the overlap integral \( {\mathcal{I}}(t_{0}) \) is defined as 
\begin{eqnarray}
{\mathcal{I}}(t_{0})=\int ^{\infty }_{-\infty }\int ^{\infty }_{-\infty }d\tau d\tau' \tanh (\tau ){\textrm{sech}}^{2}(\tau )\tanh (\tau' ){\textrm{sech}}^{2}(\tau' ){\tilde{\mathcal{F}}}(\tau/A -\tau'/A ).\label{calI} 
\end{eqnarray}
 Here \( {\tilde{\mathcal{F}}}(\tau ) \) is the inverse Fourier transform of
the fluorescence \( {\mathcal{F}}(\Omega )=\frac{1}{2}[n_{th}({\Omega })+\frac{1}{2}]\alpha ^{R}({\Omega }). \)

The correlations in position fluctuations correspond to the jitter in arrival
times, because we have chosen a propagative reference frame. The jitter therefore
feeds off position fluctuations as well as noise entering through the velocity:
\begin{eqnarray}
 &  & \langle \Delta q(\zeta )\Delta q^{*}(\zeta' )\rangle =\langle \Delta q(0)\Delta q^{*}(0)\rangle \nonumber \\
 & + & \int _{0}^{\zeta}\int _{0}^{\zeta'}d\zeta'' d\zeta''' \left[ A^{2}\langle \Delta V(\zeta'' )\Delta V^{*}(\zeta''' )\rangle +\langle \Gamma _{q}(\zeta'' )\Gamma _{q}^{*}(\zeta''' )\rangle \right] \quad \quad \zeta <\zeta' .
\end{eqnarray}
 Thus the timing jitter is 
\begin{eqnarray}
\langle [\Delta \tau (\zeta )]^{2}\rangle  & = & \langle \Delta q(\zeta )\Delta q^{*}(\zeta )\rangle \nonumber \\
 & = & \frac{\pi ^{2}}{24\overline{n}}+\frac{\pi ^{2}\alpha ^{G}}{12\overline{n}}\zeta +\frac{A^3}{6\overline{n}}\zeta ^{2}+\left[ \frac{\alpha ^{G}A^3}{9\overline{n}}+\frac{2A^4{\mathcal{I}}(t_{0})}{3\overline{n}}\right] \zeta ^{3},\label{jtr} 
\end{eqnarray}
 which contains cubic terms due to the gain and Raman couplings, and also slower
growing terms due to the initial vacuum fluctuations and amplifier noise.

We note that an alternative method that exploits conserved quantities in the NLS equation is often used\cite{s35,s65,s83} for deriving the
timing jitter.  The linearised approach that we have presented has the advantage that derivatives of products of stochastic variables do not appear.  With such derivatives, the normal rules of calculus do not apply.  Rather, the rules of Ito stochastic calculus must be observed, leading to extra drift terms. 
\subsection{Dark solitons}

Fibers in the normal dispersion regime can support dark soliton solutions, so
called since they correspond to a dip in the background intensity\cite{s50}:
\begin{eqnarray}
\phi _{\textrm{dark}}(\tau ,\zeta ) & = & \phi _{0}\sqrt{1-A^{2}{\textrm{sech}}^{2}[\phi _{0}A\tau -q(\zeta )]}\exp[i\theta (\zeta )]\exp[i\sigma (\zeta ,\tau )],\nonumber \\
\sigma (\zeta ,\tau ) & = & \arcsin \left\{ \frac{A\tanh [\phi _{0}A\tau -q(\zeta )]}{\sqrt{1-A^{2}{\textrm{sech}}^{2}[\phi _{0}A\tau -q(\zeta )]}}\right\} ,
\end{eqnarray}
 where \( d\theta /d\zeta =\phi _{0}^{2} \), \( dq/d\zeta =A\sqrt{1-A^{2}}\phi _{0}^{2} \)
and \( \phi _{0} \) is the amplitude of the background field. The size of the intensity
dip at the center of the soliton is given by \( A \), with the intensity going
to zero in a black \( \tanh (\tau ) \) soliton, for which \( A=1 \). Dark
solitons are classed as topological solitons, because they connect two background
pulses of different phase. The total phase difference between the boundaries
is \( \psi =2\arcsin (A) \).

The nonvanishing boundary conditions of the dark pulse complicate the perturbation calculation of jitter variance.  To ensure that all relevant integrals take on finite values, we impose periodic boundary conditions at $\tau = \pm \tau_l$, which are taken to infinity at the end of the calculation.  These boundary conditions require a soliton solution of the form
\begin{eqnarray}
\phi _{\textrm{dark}}(\tau ,\zeta ) & = & \phi _{0}\exp[i\theta (\zeta ) - i\kappa\tau]\left\{ \cos{\frac{\psi}{2}}+i\sin{\frac{\psi}{2}}\tanh [\phi _{0}\tau\sin{\frac{\psi}{2}} -q(\zeta )]\right\} ,
\end{eqnarray}
with a wavenumber offset $\kappa = \frac{1}{\tau_l}\arctan\left[ \tan{\frac{\psi}{2}}\tanh(\phi_0\tau_l\sin{\frac{\psi}{2}})\right]$.  The perturbation theory now proceeds in a similar fashion to the bright soliton case, except that we can greatly simplify the calculation if the unperturbed solution is taken to be a black soliton, i.e. $\psi = \pi$.  

The projection functions for the soliton parameters $P_{i}\in \{\psi,q,\phi_0,\theta \}$ are 
\begin{eqnarray}
f_{\theta} & = & -\phi_0\tanh(\phi_0 \tau-q)\exp(i\theta-i\kappa\tau)\nonumber \\
f_{\phi_0} & = & i[\tanh(\phi_0 \tau-q) +\phi_0\tau{\textrm{sech}}^2(\phi_0 \tau-q)]\exp(i\theta-i\kappa\tau)\nonumber \\
f_{q} & = & -i\phi_0{\textrm{sech}}^2(\phi_0 \tau-q)\exp(i\theta-i\kappa\tau)\nonumber \\
f_{\psi} & = & \phi_0[\beta_1\phi_0 \tau\tanh(\phi_0 \tau-q) - \frac{1}{2}]\exp(i\theta-i\kappa\tau),
\end{eqnarray}
where $\beta_1 = 1/[2\phi_0\tau_l\tanh(\phi_0\tau_l)]$.  For the required adjoint functions, we choose:
\begin{eqnarray}
\underline{f_{q}} & = & \frac{-i3\gamma_q}{4}{\textrm{sech}}^2(\phi_0 \tau-q)\exp(i\theta-i\kappa\tau)\nonumber \\
\underline{f_{\psi}} & = & \frac{\gamma_{\psi}}{\beta_1-1}{\textrm{sech}}^2(\phi_0 \tau-q)\exp(i\theta-i\kappa\tau),
\end{eqnarray}
where $\gamma_q = 4/(3\int^{\phi_0\tau_l}_{-\phi_0\tau_l}dt {\textrm{sech}}^4t)$ and $\gamma_{\psi} = (\beta_1-1)/\int^{\phi_0\tau_l}_{-\phi_0\tau_l}dt  (\beta_1 t\tanh{t}{\textrm{sech}}^2{t} - {\textrm{sech}}^2{t}/2)$.  The orthogonality condition is now
\begin{eqnarray}
\Re \left\{ \int _{-\tau_l +q/\phi_0}^{\tau_l +q/\phi_0}d\tau f_{P_{i}}\underline{f_{P_{j}}}^{*}\right\} =\delta _{i,j}.
\end{eqnarray}

Once again, the adjoint functions can be used in the linearized equation [Eq.\ (\ref{linear})] to determine how the fluctuations in position evolve:
\begin{eqnarray}
\frac{\partial }{\partial \zeta }\Delta q(\zeta ) &=& \frac{\phi_0\beta_2}{2}\Delta \psi(\zeta ) + \Gamma_q(\zeta ) \nonumber \\
\frac{\partial }{\partial \zeta }\Delta \psi(\zeta ) &=& \Gamma_{\psi}(\zeta ),
\end{eqnarray}
where $\beta_2 = \phi_0\tanh(\phi_0 \tau_l) - 1/\tau_l$.  Here we see how the fluctuations in phase produce fluctuations in position.  The stochastic term in the equation for $\psi$ evaluates to 
\begin{eqnarray}
\Gamma _{\psi}(\zeta ) & = & \Re \left\{ \int _{-\tau_l +q/\phi_0 }^{\tau_l +q/\phi_0 }d\tau \underline{f_{\psi}}^{*}(\zeta )\overline{\Gamma }(\tau ,\zeta )\right\} \nonumber \\
 & = & \int _{-\tau_l +q/\phi_0 }^{\tau_l +q/\phi_0}d\tau \frac{\gamma_{\psi}}{\beta_1-1}{\textrm{sech}}^2(\phi_0 \tau -q)\left[ \Re \{\exp(-i\theta+i\kappa\tau)\Gamma \} - \phi_0 \tanh(\phi_0\tau -q)\Gamma ^{R}\right],
\end{eqnarray}
from which the correlations of the phase fluctuations can be calculated: 
\begin{eqnarray}
\langle \Delta \psi(\zeta )\Delta \psi^{*}(\zeta' )\rangle  & = & \bigl <{\Delta \psi(0)\Delta \psi^{*}(0)}\bigr >+\int _{0}^{\zeta }\int _{0}^{\zeta' }d\zeta''d \zeta''' \langle \Gamma _{\psi}(\zeta'' )\Gamma _{\psi}^{*}(\zeta''' )\rangle \nonumber \\
 & = & \frac{{\gamma_\psi}^2}{3\overline{n}\gamma_q\phi_0}+\left[ \frac{2\alpha ^{G}{\gamma_\psi}^2}{3\overline{n}\gamma_q(\beta_1-1)^2\phi_0}+\frac{2{\gamma_\psi}^2{\mathcal{I}}_{\tau_l}(t_{0})}{\overline{n}(\beta_1-1)^2}\right] \zeta \quad \quad \zeta <\zeta' ,
\end{eqnarray}
 where the overlap integral \( {\mathcal{I}}_{\tau_l}(t_{0}) \) is now defined as 
\begin{eqnarray}
{\mathcal{I}}_{\tau_l}(t_{0})=\int ^{\phi_0\tau_l}_{-\phi_0\tau_l }\int ^{\phi_0\tau_l}_{-\phi_0\tau_l }d\tau d\tau' \tanh (\tau ){\textrm{sech}}^{2}(\tau )\tanh (\tau' ){\textrm{sech}}^{2}(\tau' ){\tilde{\mathcal{F}}}(\tau/\phi_0 -\tau'/\phi_0 ). 
\end{eqnarray}

The leading order terms for the fluctuations in position are thus
\begin{eqnarray}
\langle \Delta q(\zeta )\Delta q^{*}(\zeta)\rangle &=& \frac{\phi_0^2\beta_2^2}{4}\int _{0}^{\zeta}\int _{0}^{\zeta'}d\zeta'' d\zeta''' \langle \Delta \psi(\zeta'' )\Delta \psi^{*}(\zeta''' )\rangle \nonumber \\
&=& \frac{\phi_0\beta_2^2{\gamma_{\psi}}^2}{12 \overline{n}\gamma_q}z^2 + \left[ \frac{\alpha ^{G}\phi_0\beta_2^2{\gamma_{\psi}}^2}{18\overline{n}\gamma_q(\beta_1-1)^2} +\frac{{\mathcal{I}}_{\tau_l}(t_{0})\phi_0^2\beta_2^2{\gamma_{\psi}}^2}{6\overline{n}}\right] \zeta ^{3}.
\end{eqnarray}
By taking the limit $\tau_l \rightarrow \infty$, we find the leading order terms in the jitter growth for a black soliton to be
\begin{eqnarray}
\langle [\Delta \tau (\zeta )]^{2}\rangle =\frac{\phi_0^3}{12\overline{n}}\zeta ^{2}+\left[ \frac{\alpha ^{G}\phi_0^3}{18\overline{n}}+\frac{{\mathcal{I}}(t_{0})\phi_0^4}{6\overline{n}}\right] \zeta ^{3},
\end{eqnarray}
where the overlap integral \( {\mathcal{I}}(t_{0}) \) is as defined in Eq.\ (\ref{calI}).
As in the anomalous dispersion regime, the vacuum fluctuations contribute to
quadratic growth in the jitter variance, and gain and Raman fluctuations contribute
to cubic growth. However, the size of the jitter is smaller than that in the
bright soliton case, for the same propagation distance \( \zeta  \). The contribution
from the vacuum and gain terms is one half and the contribution from the Raman
term is one quarter of that in Eq.\ (\ref{jtr}), giving dark solitons some
advantage over their bright cousins. \label{dksol}

\section{Scaling properties}

\label{SP} In summary, there are three different sources of noise in the soliton,
all of which must be taken into account for small pulse widths. These noise
sources contribute to fluctuations in the velocity parameter, which lead to
quadratic or cubic growth in the timing-jitter variance for single-pulse propagation.
The noise sources also produce other effects, such as those effected through
soliton interactions, but we will not consider these here.

Each of the noise sources has different characteristic scaling properties, which
are summarized as follows:

\subsection{Vacuum Fluctuations}

The vacuum fluctuations cause diffusion in position which is important for small
propagation distances. There are position fluctuations even at the initial position,
since the shot noise in the arrival time of individual coherent-state photons
gives an initial fluctuation effect. After propagation has started, this initial
position fluctuation is increased by the additional variance in the soliton
velocity, due essentially to randomness in the frequency domain.

For bright solitons the resulting soliton timing variance is given by\cite{s24}
\begin{eqnarray}
\langle [\Delta \tau (\zeta )]^{2}\rangle _{I}=\frac{\pi ^{2}}{24\overline{n}}+\frac{1}{6\overline{n}}\zeta ^{2}\quad \quad {(\textrm{bright})}. & 
\end{eqnarray}
 For purposes of comparison, note that \( \overline{N}=2\overline{n} \) is
the mean photon number for a \( {\textrm{sech}}(\tau ) \) soliton. Numerical
calculations confirm that for \( \tanh (\tau ) \) dark solitons, the variance
was about one half the bright-soliton value, as predicted by the analysis in
outlined in Sec.\ \ref{dksol}: 
\begin{eqnarray}
\langle [\Delta \tau (\zeta )]^{2}\rangle _{I}=\frac{\pi ^{2}}{48\overline{n}}+\frac{1}{12\overline{n}}\zeta ^{2}\quad \quad {(\textrm{dark})}. & 
\end{eqnarray}
 This shot-noise effect, which occurs without amplification, is simply due to
the initial quantum-mechanical uncertainty in the position and momentum of the
soliton. Because of the Heisenberg uncertainty principle, the soliton momentum
and position cannot be specified exactly. This effect dominates the Gordon-Haus
effect over propagation distances less than a gain length. However, for short
pulses, this distance can still correspond to many dispersion lengths - thus
generating large position jitter.  We note that there are also initial fluctuations
in the background continuum, which may feed into the soliton parameters as the soliton 
propagates.  This lessor effect is included in the numerical 
calculation; a comparison of the numerical results with the analytic results 
confirms that the initial fluctuations in the soliton parameters account for almost all
of the shot-noise contribution to the jitter.

\subsection{Gordon-Haus noise}

As is well known, the noise due to gain and loss in the fiber produces the Gordon-Haus
effect, which is currently considered the major limiting factor in any long-distance
soliton-based communications system using relatively long (\( >10ps \)) pulses.
Amplification with mean intensity gain \( \alpha ^{G} \), chosen to compensate
fiber loss, produces a diffusion (or jitter) in position. Unless other measures
are taken, for sufficiently small amplifier spacing\cite{e375} and at large
distances this is given by\cite{s35,s65,s45}
\begin{eqnarray}
\langle [\Delta \tau (\zeta )]^{2}\rangle _{GH} & \simeq  & \frac{\alpha ^{G}}{9\overline{n}}\zeta ^{3}\quad \quad {(\textrm{bright})},\nonumber \\
\langle [\Delta \tau (\zeta )]^{2}\rangle _{GH} & \simeq  & \frac{\alpha ^{G}}{18\overline{n}}\zeta ^{3}\quad \quad {(\textrm{dark})},
\end{eqnarray}
 in which the linearly growing terms have been neglected.

Another effect of the amplifier noise is to introduce an extra noise term via
the fluctuations in the Raman-induced soliton self-frequency shift. This term
scales as the fifth power of distance and hence will become important for long
propagation distances. This combined effect of spontaneous emission noise and
the Raman intrapulse scattering has been dealt with by others\cite{MooWonHau94}.
The full phase-space equation [Eq. (\ref{RMS})] models this accurately, since it includes the delayed
Raman nonlinearity, and the effect would be seen in numerical simulations
carried out over long propagation distances.

\subsection{Raman noise}

A lesser known effect are the fluctuations in velocity that arise from the Raman
phase-noise term \( \Gamma ^{R} \) in Eq.\ (\ref{NLSE}). Like the Gordon-Haus
effect, this Raman noise generates a cubic growth in jitter variance: 
\begin{eqnarray}
\langle [\Delta \tau (\zeta )]^{2}\rangle _{R} & = & \frac{2{\mathcal{I}}(t_{0})}{3\overline{n}}\zeta ^{3}\quad \quad {(\textrm{bright})},\nonumber \\
\langle [\Delta \tau (\zeta )]^{2}\rangle _{R} & = & \frac{{\mathcal{I}}(t_{0})}{6\overline{n}}\zeta ^{3}\quad \quad {(\textrm{dark})}.
\end{eqnarray}
 where \( {\mathcal{I}}(t_{0}) \) is the integral defined in Eq.\ (\ref{calI})
that indicates the spectral overlap between the pulse spectrum and the Raman
fluorescence. The mean-square Raman induced timing jitter has a cubic growth
in both cases, but the dark soliton variance is one quarter of that of the bright
soliton.

The magnitude of this Raman jitter can be found by evaluating \( {\mathcal{I}}(t_{0}) \)
numerically, or else using an analytic approximation. An accurate model of the
Raman gain, on which \( {\mathcal{I}}(t_{0}) \) depends, requires a multi-Lorentzian
fit to the experimentally measured spectrum\cite{RamanGain}. A fit with 11 Lorentzians was used
in the numerical simulations, including 10 Lorentzians to accurately model the measured
gain and fluorescence. One extra Lorentzian was used at low frequencies to model GAWBS
(Guided Wave Acoustic Brillouin Scattering); this has a relatively small effect on
an isolated soliton, except to cause phase noise.

For analytic work, however, a single-Lorentzian model\cite{s76} can
suffice for approximate calculations.  A  plot of the Raman gain profile  $\alpha ^R(\Omega )$ for both models is given in (QNI), along with a table of the fitting parameters for the multi-Lorentzian model.  The spectral features of the Raman noise correlations are determined directly from the Raman fluorescence function ${\mathcal{F}}(\Omega )$, which we plot in Fig.\ \ref{solspect}.  For the single-Lorentzian model, the fluorescence spectrum is approximately flat at low frequencies: 
\begin{eqnarray}
{\mathcal{F}}(\Omega ) & = & \frac{1}{2}\left[ n_{th}(\Omega )+\frac{1}{2}\right] \alpha ^R(\Omega )\nonumber \\
 & \simeq  & \frac{2F_{1}\Omega _{1}\delta _{1}^{2}k_{B}T}{(\Omega _{1}^{2}+\delta _{1}^{2})^{2}\hbar }={\mathcal{F}}(0),
\end{eqnarray}
 which greatly simplifies the Raman correlations. As Fig.\ \ref{solspect}
indicates, the spectral overlap of \( {\mathcal{F}}(\Omega ) \) with a \( t_{0}=1ps \)
soliton occurs in this low frequency region. Thus the white noise approximation
for the Raman correlations is good for solitons of this pulse width and larger.
For smaller pulse widths, not only is the Raman contribution to the noise larger
due to the greater overlap, but the colored nature of the correlations must
be taken into account.

In the single-Lorentzian model, \( {\mathcal{I}}(t_{0}\rightarrow \infty )\simeq \frac{4}{15}{\mathcal{F}}(0) \),
which gives \label{sol_approx}
\begin{eqnarray}
{\langle [\Delta t(x)]^{2}\rangle }_{R} & \simeq  & \frac{8|k''|^{2}n_{2}\hbar \omega_{0}^{2}{\mathcal{F}}(0)}{45{\cal A}ct_{0}^{3}}x^{3} = \frac{8t_{0}^{2}{\mathcal{F}}(0)}{45{\overline{n}}}\left( \frac{x}{x_{0}}\right) ^{3}\quad \quad {(\textrm{bright})},\nonumber \\
{\langle [\Delta t(x)]^{2}\rangle }_{R} & \simeq  & \frac{2|k''|^{2}n_{2}\hbar \omega_{0}^{2}{\mathcal{F}}(0)}{45{\cal A}ct_{0}^{3}}x^{3} = \frac{2t_{0}^{2}{\mathcal{F}}(0)}{45{\overline{n}}}\left( \frac{x}{x_{0}}\right) ^{3}\quad \quad {(\textrm{dark})}.
\end{eqnarray}
 At a temperature of \( 300K \), \( {\mathcal{F}}(0)=4.6\times10 ^{-2} \)
when a single Lorentzian centered at 12 THz with fitting parameters \( F_{1}=0.7263 \),
\( \delta _{1}=20\times10 ^{12}t_{0} \) and \( \Omega _{1}=75.4\times10 ^{12}t_{0} \)
is used.

\section{Numerical results}

\label{NR} \label{numres} More precise results can be obtained by numerically integrating the original Wigner phase-space equation, Eq. (\ref{RMS}), which includes the full time-delayed nonlinear Raman response function.  The results for \( t_{0}=500fs \)
bright and dark solitons are shown in Figs.\ \ref{ol}(a) and \ref{ol}(b),
respectively. The gain and photon number were chosen to be \( G=\alpha ^{G}/x_{0}=4.6\times {10}^{-5}m^{-1}\, (0.2dB/km) \)
and \( \overline{n}=4\times 10^{6} \), with \( x_{0}=440m \). These values
are based on \( {\cal A}=40[\mu m]^{2} \), \( k''=0.57[ps]^{2}/km \), and \( n_{2}=2.6\times10 ^{-20}[m]^{2}/W \)
for a dispersion-shifted fiber\cite{e34}. These numerical calculations use
the multiple-Lorentzian model of the Raman response function shown in Fig.\ \ref{solspect},
which accurately represents the detailed experimental response function.  

The numerical method is based on the split-step idea\cite{Drummond1983}
as adapted to Raman propagation\cite{s23}. Noise is treated using a central
difference technique appropriate to stochastic equations\cite{Drummond1991},
with the necessary adaptations required to treat a partial stochastic differential
equation \cite{Werner1997}. All calculations were duplicated using two different
space steps, but with the same underlying noise sources, in order to calculate
discretization error. Sampling error was also estimated using standard central
limit theorem procedure over a large ensemble of noise sources. Tests on time steps
and window sizes were also carried out to ensure there were no errors from these
sources.

The initial conditions consist of a coherent laser pulse injected into the fiber.
In the Wigner representation, this minimum uncertainty state leads to the initial
vacuum fluctuations. The numerical calculation thereby includes the full effect of
these zero-point fluctuations, including the noise that appears in the 
background continuum and in the soliton parameters.  We take the initial 
pulse shape in the anomalous dispersion
regime to be a fundamental bright soliton, with \( A=1 \) and \( V=\theta =q=0 \).
Such a soliton can be experimentally realized with a sufficiently intense pulse,
which will reshape into a soliton or soliton train. The nonsoliton part of
the wave will disperse and any extra solitons will move away at different velocities
from the fundamental soliton. The numerical simulation in the normal dispersion
regime used two black solitons of opposite phase chirp (\( A=\pm1  \)), so
that the field amplitude at either boundary was the same. This phase matching
ensured the stability of the numerical algorithm, which assumes periodic boundary
conditions.

The position jitter at a given propagation time was calculated by combining
the waveform with a phase-matched local-oscillator pulse that had a linear chirp
in amplitude, and integrating the result to give the soliton position. This
homodyne measurement involves symmetrically ordered products, so the Wigner
representation will give the correct statistics. The variance in soliton position
was then calculated from a sample of 1000 trajectories. For this small distance
of propagation (\( \simeq 10km \)), the jitter variance due to the initial
noise is twice the Gordon-Haus jitter, but for larger distances, the cubic effects
are expected to dominate.

For ultrafast pulses, the Raman jitter dominates the Gordon-Haus jitter (by
a factor of two in the \( 500fs \) bright soliton case) and will continue to
do so even for long propagation distances. For short propagation distances,
the Gordon-Haus effect is not exactly cubic, because of neglected terms in the
perturbation expansion, which give a linear (as opposed to cubic) growth in
the jitter variance. However, there are no such terms in the Raman case. The
analytic Raman results are also shown in the figures, which show that our approximate
formula gives a reasonable fit to the numerical data even for subpicosecond
pulses. Using this approximate formula, the relative size of the two effects
scales as 
\begin{eqnarray}
\frac{{\langle [\Delta t(x)]^{2}\rangle }_{R}}{{\langle [\Delta t(x)]^{2}\rangle }_{GH}} & = & \frac{6{\mathcal{I}}(t_{0})|k''|}{G{t_{0}}^{2}}=\frac{6{\mathcal{I}}(t_{0})}{Gx_{0}}\simeq \frac{8{\mathcal{F}}(0)}{5Gx_{0}}\quad \quad {(\textrm{bright})},\nonumber \\
\frac{{\langle [\Delta t(x)]^{2}\rangle }_{R}}{{\langle [\Delta t(x)]^{2}\rangle }_{GH}} & = & \frac{3{\mathcal{I}}(t_{0})|k''|}{G{t_{0}}^{2}}=\frac{3{\mathcal{I}}(t_{0})}{Gx_{0}}\simeq \frac{4{\mathcal{F}}(0)}{5Gx_{0}}\quad \quad {(\textrm{dark})},
\end{eqnarray}
 These equations show why experiments to date\cite{e34}, which have used longer
pulses (\( t_{0}>1ps \)) and dispersion-shifted fiber, have not detected the
Raman-noise contribution to the jitter. The Raman jitter exceeds the Gordon-Haus
jitter for bright solitons with periods \( x_{0}<1.5km \). Dark solitons, on
the other hand, have an enhanced resistance to the Raman noise, which means
that a shorter period is needed before the Raman jitter will become important.

The total jitter, which corresponds to the realistic case where all three noise
sources are active, is also shown in Fig.\ \ref{ol}, and, in the bright soliton
case, is about a factor of three larger than the ordinary Gordon-Haus effect,
over the propagation distance shown. The physical origin of these quantum noise
sources cannot easily be suppressed. The initial vacuum-induced timing jitter
is caused by the shot-noise variance in the soliton guiding frequency. The physical
origin of the Raman jitter is in frequency shifts due to soliton phase modulation
by the ever-present quantum and thermal phonon fields in the fiber medium.

The numerical method should give accurate results far beyond the distance shown in 
Fig. \ref{ol}, provided that the transverse and propagative resolutions are made large 
enough.  The equations generated in the Wigner method should remain valid up to $\zeta 
\sim \sqrt{\overline{n}}$, which corresponds to about $1000 km$.  When the Wigner equations
can no longer be trusted, the positive-$P$ equations will still give accurate results.  This 
paper has not analysed any multisoliton effects, although the numerical method does simulate 
interactions between solitons.  This just requires the initial conditions and simulation window width to 
be set up according to whether interactions are to be considered or not.  We 
have not included third-order dispersion in our model, which would become important for very 
small pulses ($t_0 \simeq 100 fs$), but it could easily be included into the equations for numerical simulation.

The approximate analytic results are most limited probably by their exclusion of Raman intrapulse effects, such as
the deterministic self-frequency shift and the amplifier jitter that feeds through this.  Approximate calculations\cite{MooWonHau94} of the self-frequency shift jitter variance show that it grows as the fifth power of distance.  With our parameters and the measured value of the Raman time constant\cite{AtiMysChrGal}, it would become larger than the usual Gordon-Haus effect at $x \simeq 100 km$, or about 10 times the propagation distance shown in Fig. \ref{ol}.  Using standard techniques\cite{MooWonHau94}, the perturbation theory presented in this paper could be extended to include the self-frequency shift contributions (from both the amplifier noise and Raman phase noise) to the total jitter.

\section{Conclusions}

\label{Cs} Our major conclusion is that quantum noise effects due to the intrinsic
finite-temperature phonon reservoirs are a dominant source of fluctuations in
phase and arrival time, for subpicosecond solitons. For longer solitons, Raman
effects are reduced when compared to the Gordon-Haus jitter from the laser gain
medium that is needed to compensate for losses. The reason for this is the smaller
intensity of the pulse, and therefore the reduced Raman couplings that occur
for longer solitons -- which are less intense than shorter solitons with the
same dispersion. The ratio can be calculated simply from the product \( Gx_{0} \),
which gives the gain per soliton length. A smaller \( x_{0} \) corresponds
to a shorter, more intense soliton and hence a larger Raman noise; while a larger
\( G \) corresponds to increased laser gain, with larger spontaneous noise.

At a given pulse duration and fiber length, a strategy for testing this prediction
would be to use short pulses with dispersion-shifted fiber having an increased
dispersion, since this increases the relative size of the Raman jitter. The
physical reason for this is very simple. Solitons have an intensity which increases
with dispersion if everything else is unchanged. At the same time, the multiplicative
phase noise found in Raman propagation is proportional to intensity, and hence
becomes relatively large compared to the additive Gordon-Haus noise due to amplification.
For large enough dispersion, the temperature-dependent Raman jitter should become
readily observable at short enough distances so that amplification is unnecessary.
This would give a completely unambiguous signature of the effect we have calculated.
A mode-locked fiber soliton laser would be a suitable pulse source, owing to
the very short (\( 64fs \)), quiet pulses\cite{Yu1997,Namiki1996} that are
obtainable.

\acknowledgments

We would like to acknowledge helpful comments on this paper by Wai S. Man.

\begin{figure}
\caption{Spectrum of the fluorescence function \protect\protect\( {\mathcal{F}}(\omega )\protect 
\protect \) for the 11-Lorentzian model (continuous lines) and the single-Lorentzian model
(dashed lines), for a temperature of \protect\protect\( T=300K\protect \protect \).
Also shown is the spectrum of a \protect\protect\( t_{0}=1ps\protect \protect \)
soliton.}

\label{solspect}
\end{figure}

\begin{figure}
\caption{Timing jitter in \protect\protect\( t_{0}=500fs\protect \protect \) bright
(a) and dark (b) solitons due to initial quantum fluctuations (circles), Gordon-Haus
effect (crosses) and Raman noise (plus signs). The asterisks give the total
jitter and the continuous line gives the approximate analytic results for the
Raman jitter. }

\label{ol}
\end{figure}

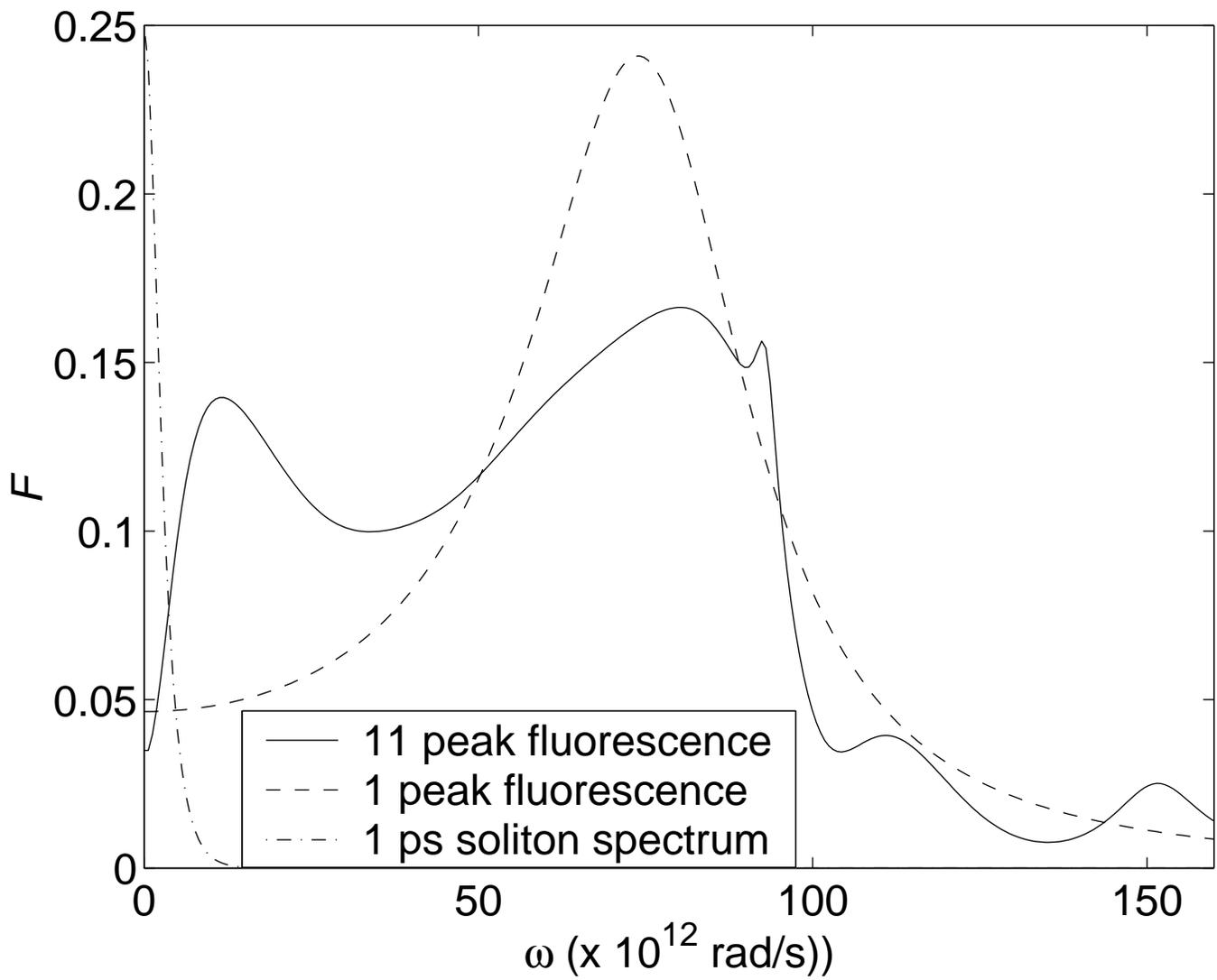
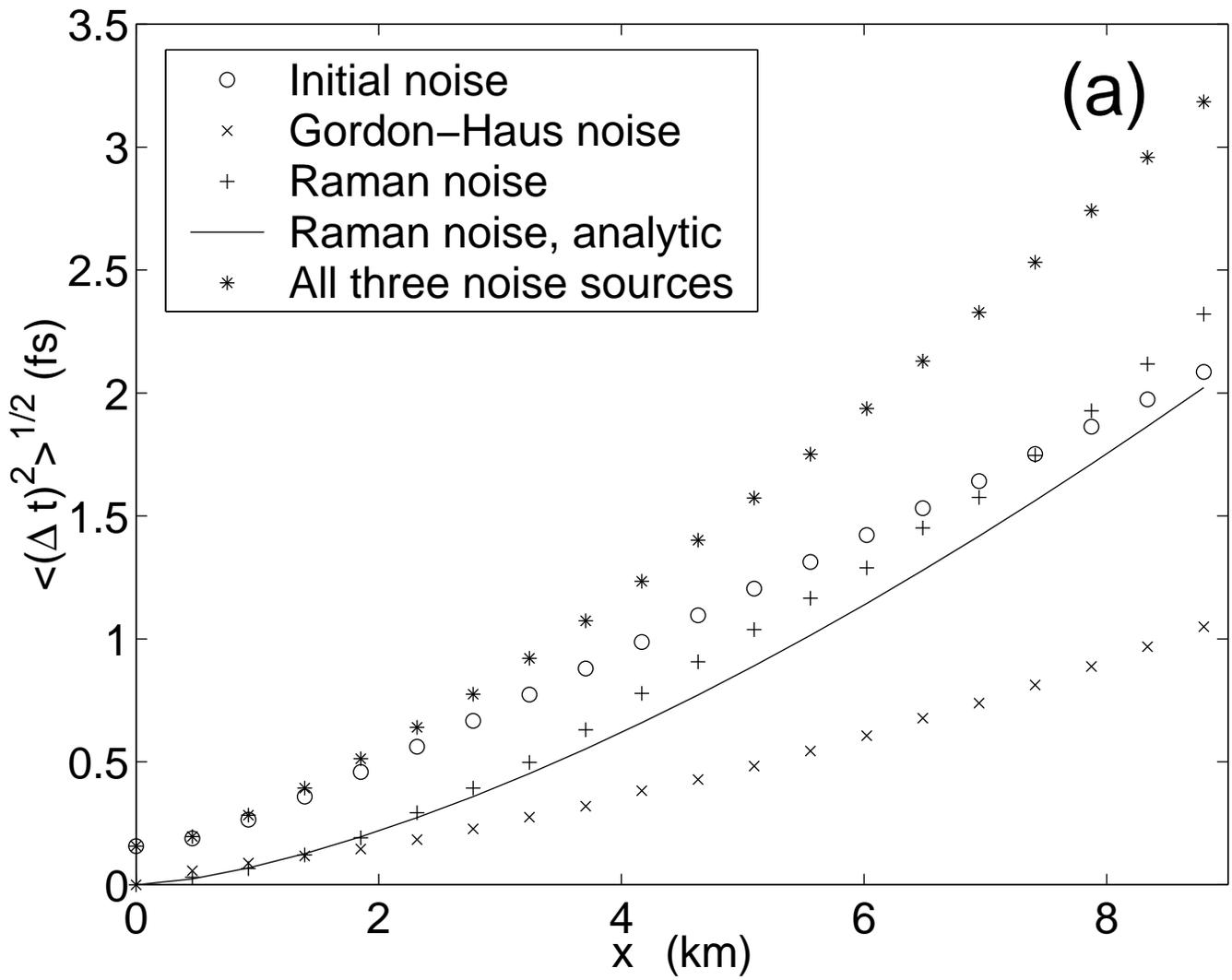
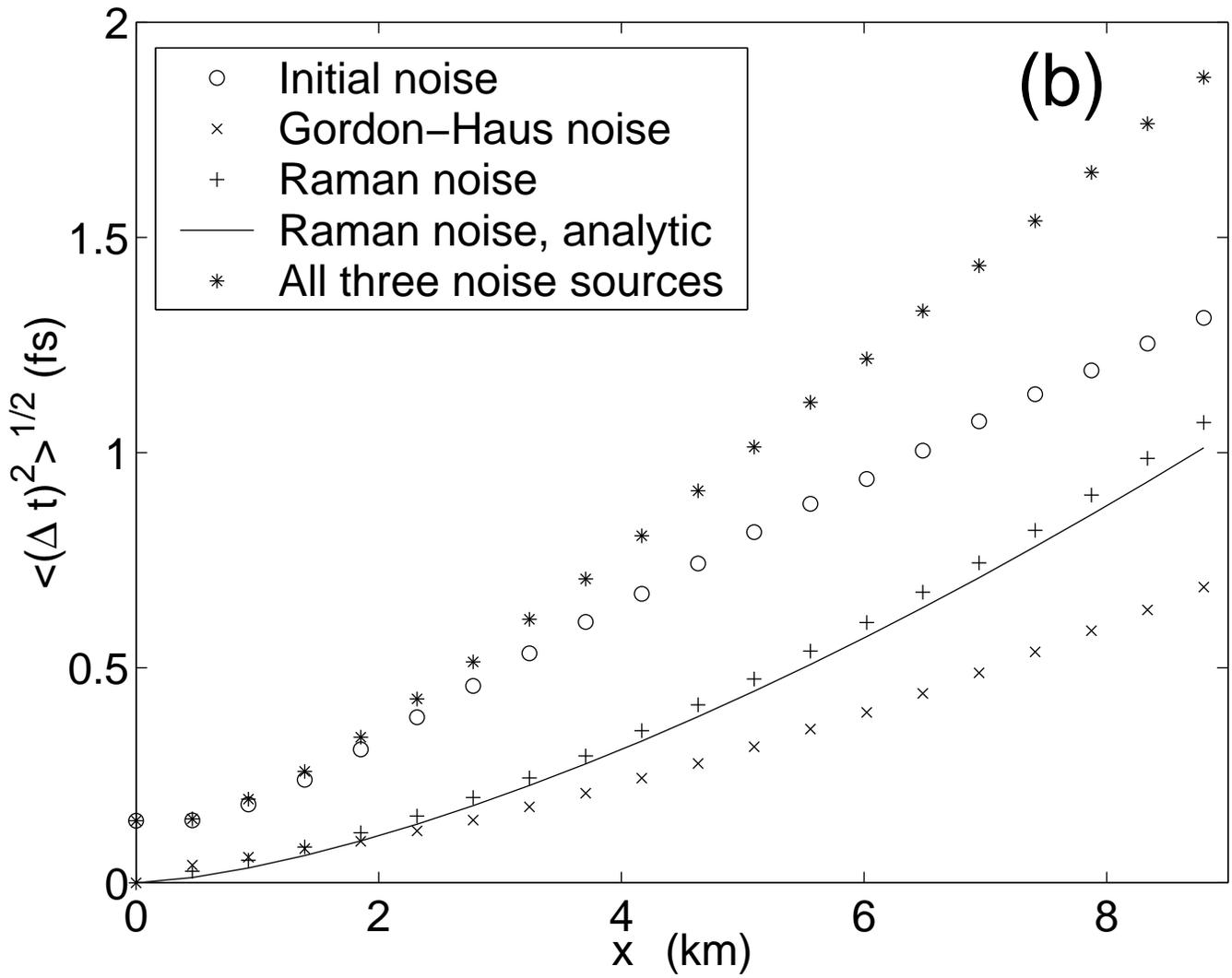
\end{document}